\documentclass[12pt]{iopart}

\usepackage{siunitx}
\usepackage{graphicx}
\usepackage{cite}
\usepackage{xcolor}

\begin{document}

\title[\textit{N}-Carbophenes]{\textit{N}-Carbophenes: two-dimensional covalent organic frameworks derived from linear \textit{N}-phenylenes}

\author{Chad E. Junkermeier$^1$, Jay Paul Luben$^1$, Ricardo Paupitz$^2$}
\address{$^1$Department of Science, Technology, Engineering, and Mathematics, University of Hawai`i Maui College, Kahului HI 96732, USA}

\address{$^2$Departamento de F\'{\i}sica, IGCE, Universidade Estadual Paulista, UNESP, 13506-900, Rio Claro, SP, Brazil}
\ead{junkerme@hawaii.edu}

\vspace{10pt}

\begin{abstract}
\textit{N}-Carbophene (carbophene) is a novel class of two-dimensional covalent organic frameworks (2DCOF), based on linear \textit{N}-phenylenes, that have moderate band gaps and low-mobility bands surrounding the Fermi energy; the simplest of which may have been recently synthesized.  Using tight-binding density functional theory, the ground state configurations single layers, bilayers, and bulk systems was determined.  This work finds that carbophenes have formation energies per carbon atom similar to that of graphenylene.  The similarity of formation energies between graphenylene and carbophene suggests that when trying to synthesize one, the other may also be synthesized.  The formation energies could explain why the first reported synthesis of graphenylene also indicated that they may have synthesized 3-carbophene.  Results contained in this work suggests that a carbophene was synthesized instead of graphenylene. The projected density of states (PDOS) demonstrates that the anti-aromatic nature of the cyclobutene units plays a direct role in the creation of bands around the Fermi level, making this an exciting material in the theoretical understanding of the nature of aromatic bonds.
\end{abstract}

%
\vspace{2pc}
\submitto{\TDM}
%
%
%

\section{Introduction}
Almost from the start of the first description of a macromolecule the dream of 2DCOF has lingered in the imaginations of chemists.\cite{Staudinger19201073,flory1953principles} C{\^o}t{\'e} \textit{et al.} were the first to produce a 2DCOF; which they synthesized by employing a bottom up approach.\cite{Cote20051166}  The architecture of these structures is similar to that of graphene, but with trifunctional molecules in the place of the carbon atoms and linking molecules in place of the C-C bonds.  Methods developed by teams led by Schl\"uter decreased the amount of cross-linking between layers.\cite{Servalli2018fishnet} By changing the trifunctional and linking molecules the structures could be tailored to have specific qualities; for example, defined pore size, functionalizability, or electronic properties.\cite{Servalli2018fishnet}  The high tunability coupled with high surface area leads to materials applications such as
chemical sensing,\cite{Ding5b10754} 
drug delivery,\cite{Mitra7b00925} 
energy storage,\cite{Wang7b02648} 
gas adsorption,\cite{C7CE00344G} and
molecular separation.\cite{RajuC7NR07963J}  While many possible 2DCOFs have been discussed in the literature, some of the simplest hydrocarbons are porous 
graphene,\cite{bieri2009porous}  
graphyne, graphdiene,\cite{Psofogiannakis201219211} and 
graphbutane.\cite{Psofogiannakis201219211}

Because of their importance in developing a proper understanding of the aromaticity criterion, interest in so-called \textit{N}-phenylenes, alternating units of cyclobutene and cyclohexatriene,\cite{Song1338} has been around for decades,\cite{BARRON19662609} although development of synthesis methods has been slow.  
Interest in turning these molecules into 2DCOFs grew after the report of a reliable method of producing graphene.\cite{Novoselov04666} 
The first advance into phenylene based 2DCOFs came in the form of a one-dimensional strip of 2-phenylenes which may eventually be combined to form new carbon allotropes.\cite{ANGE:ANGE201309324, junkermeier2019bilayers}  More recently, Du \textit{et al.} reported the possible synthesis of the phenylene based 2DCOF, graphenylene.\cite{du1740796}  The presence of alternating bond lengths in the hexatomic rings of graphenylene suggests that the aromatic $\pi$-bonding present in most 2D carbons has been replaced with non-aromatic alternating single and double bonds.\cite{Song1338}  Though the bond lengths indicate graphenylene is non-aromatic in nature, band dispersion plots show that the electrons (or holes) in the conduction (valence) band can still be expected to be able to propagate through the system.\cite{Song1338, junkermeier2019bilayers} Some of the results that Du \textit{et al.} reported suggest that they had not synthesized graphenylene but, instead, a 2D hydrocarbon material based on linear 3-phenylene.\cite{du1740796} To the best of our knowledge, no experimental followup has determined which material Du \textit{et al.} synthesized. Nor has the 2D hydrocarbon (denoted here as 3-carbophene) been theoretically examined before now.

This work seeks to provide a theoretical basis for 3-carbophene and extend it to 4- through 10-carbophenes, based respectively on linear 4-phenylene through linear 10-phenylene.  Relaxed structures of each unit cell are used to discuss the formation energies of the carbophenes, which will be used to discuss Du's experimental results.  Band dispersion plots, PDOS, and charge distributions are used to show that the carbophenes are strongly non-aromatic, leading to valence and conduction bands that depending on the value of N may have low charge propagation. Using high throughput calculations, a determination of the optimum interlayer spacing for the carbophenes suggests that Du \textit{et al.} synthesized a carbophene.

\section{Methodology}\label{sec:methods}

Geometry optimizations of the structures were performed using the density functional based tight binding (DFTB) method, implemented in DFTB+.\cite{Elstner1998,aradi2007dftb,manzano2012} DFTB+ has near density functional theory (DFT) precision in electronic structure calculations while being significantly faster than DFT. Much of this speedup comes from the use of look-up tables (the so-called Slater-Koster files) instead of integral evaluation at run-time.
The {\sl matsci Slater-Koster} files, which are formulated to accurately describe materials science problems, were used.\cite{lukose2010reticular} Lennard-Jones potentials were used to simulate dispersion forces with the parameters taken from the Universal Force Field (UFF).\cite{Rappe1992UFF} For geometry optimization, including the cell parameters and atomic positions within the cell, a conjugate gradient algorithm was used. During geometry optimization, the lattice vector lengths were allowed to change, but the angle between them was not. An 8x8x1 Monkhorst-Pack kpoint grid was used during optimization, while during the density of state (DOS) calculations a 96x96x1 Monkhorst-Pack kpoint grid was employed.  The out-of-plane bounding box was set to 30 \AA.

These methods were utilized to produce the carbophene results discussed in this work. The methods were also employed to produce the graphenylene, porous graphene, graphdiene, and graphbutane values that were compared to carbophenes.

\section{Results and Discussion}

\begin{figure}[bt]
\centering
\includegraphics[clip, width=6cm, keepaspectratio]{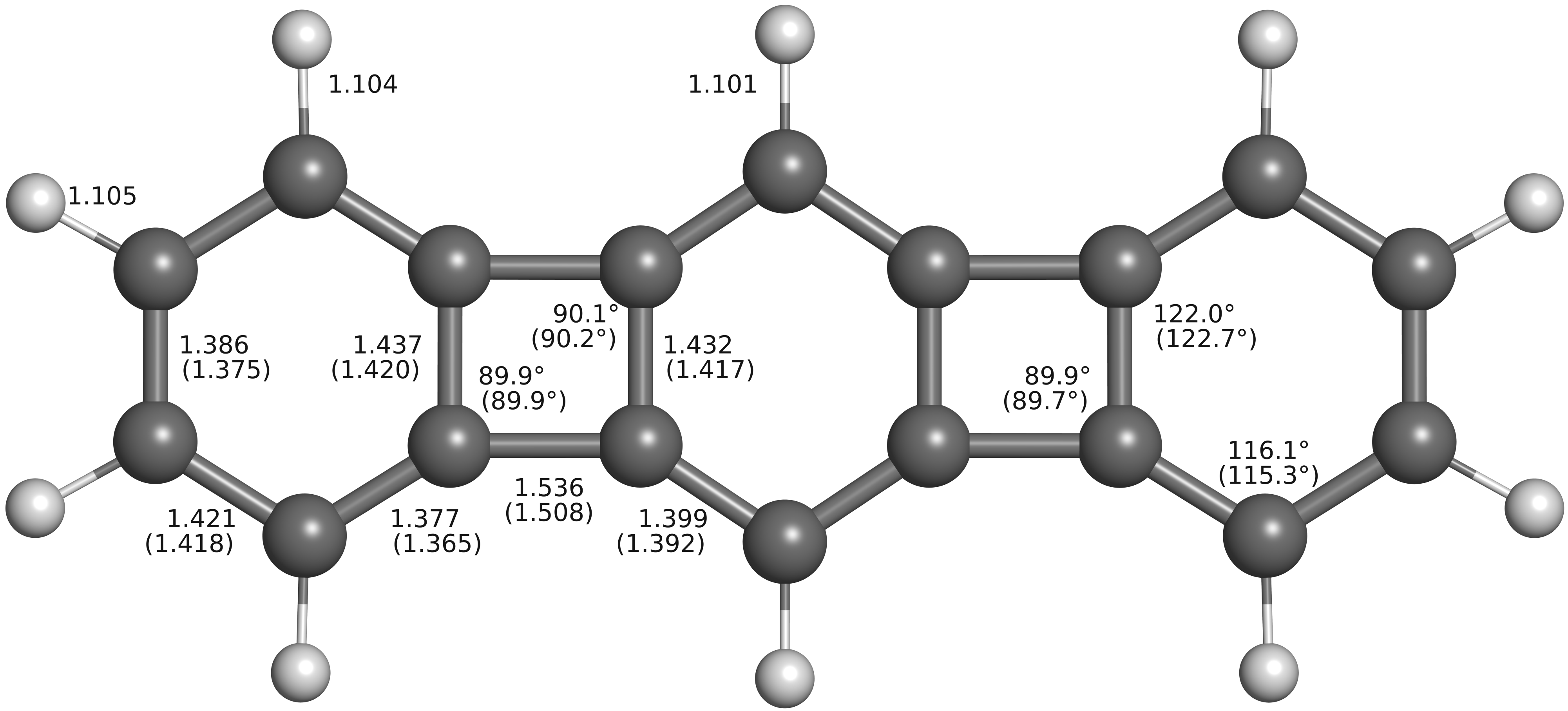}
\caption{Comparison of the bond lengths and valence bond angles in 3-phenylene as computed by DFTB+ and determined experimentally by XRD (parenthesized values).}
\label{fig:3phenylene}
\end{figure}

Previous works demonstrate that the DFTB+ method reliably reproduces the crystal structures of carbon allotropes.\cite{Brunetto1212810,junkermeier2019bilayers} The methods used here reproduced the DFT hybrid functional B3LYP band gap of graphenylene (2-carbophene).  Since hybrid functionals are known to predict band gap values to withing a few percents of the experimental value, this suggests that the method will work on the N-carbophene discussed here as well.\cite{junkermeier2019bilayers}  To further validate the results presented later, attention will first be given to 3-phenylene.  Figure \ref{fig:3phenylene} gives selected bond lengths and valence bond angles as computed using the DFTB+ method as well as the experimental values obtained from X-ray diffraction (XRD) by Schleifenbaum \textit{et al.}.\cite{SCHLEIFENBAUM20017329}  When compared, the theoretical values are within 1\% of the experimental values.  Not shown is that experimental out-of-plane dihedral angles are $< \ang{0.7}$, which fits with the planar molecule found elsewhere.\cite{SCHLEIFENBAUM20017329} Because XRD often doesn't account for the position of hydrogen atoms, experimental C-H bond lengths are not available.

\begin{figure}[bt]
\centering
\includegraphics[clip, width=6cm, keepaspectratio]{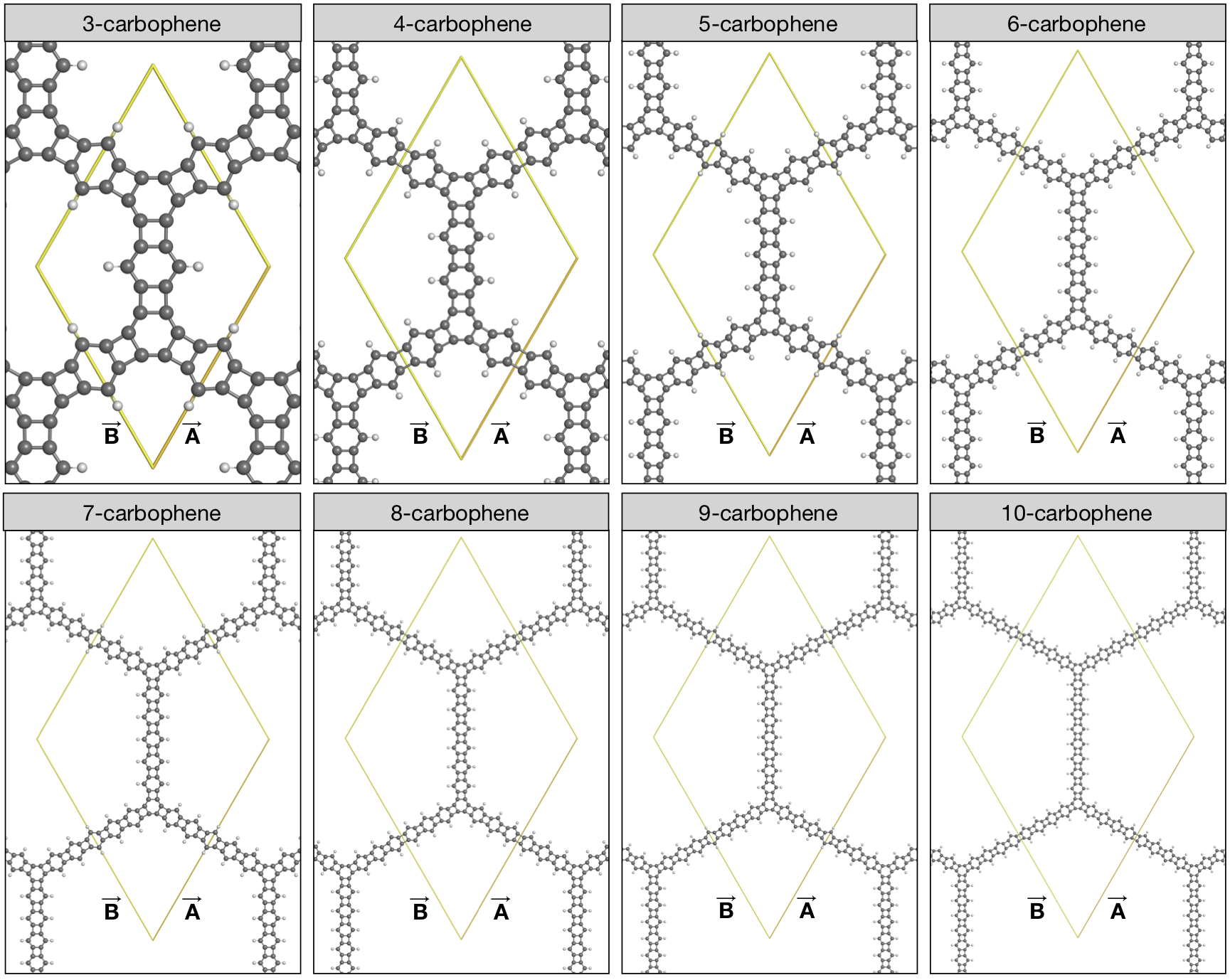}
\caption{Cell structures of carbophene systems studied.}
\label{fig:organiccells}
\end{figure}

The simplest versions of the carbophenes have hydrogen bound to each adsorption site as shown in Figure \ref{fig:organiccells}.  Table \ref{table:energylengths} gives the cell parameters, pore sizes, and formation energies of each carbophene.  The formation energies are calculated using graphene as the reference, thus avoiding errors caused by incorrectly calculating the van der Waals interactions when graphite is used as the reference.\cite{Song1338}  The formation energy changes by slightly more than 10 eV as phenylene groups are added to (N-1)-carbophene to form N-carbophene.  Interestingly, the formation energy per carbon atom for each carbophene is less than the value computed for graphenylene $0.491$ eV. The formation energy per carbon atom of the carbophenes falls between that of porous graphene $0.254$ eV and of graphdiene $0.559$ eV, and graphbutane  $0.641$ eV.

\begin{table*}[bt]
\caption{Formation energies (FE), formation energies per carbon atom (FE/C), lattice constants ($|\vec{\textit{A}}|$), H-to-H pore size (HH), C-to-C pore size (CC), sphere diameter pore size (S), bilayer interlayer spacing (BL), and bulk interlayer spacing (BK). }
\begin{scriptsize}
\begin{tabular}{ccccccccc}
\br
System & FE [eV] & FE/C [eV] & $|\vec{\textit{A}}|$ [\AA] & HH [\AA] & CC [\AA] & S [\AA] & BL [\AA] & BK [\AA] \\
\mr
    3-carbophene    	&	16.087	&	0.447	&	13.463	&	8.259	&	10.465	&	7.545	&	3.525	&	3.530 \\
    4-carbophene    	&	26.473	&	0.441	&	20.145	&	15.431	&	17.575	&	14.729	&	3.528	&	3.532 \\
    5-carbophene    	&	36.951	&	0.440	&	26.825	&	21.632	&	23.837	&	20.880	&	3.529	&	3.529 \\
    6-carbophene    	&	47.449	&	0.439	&	33.517	&	28.586	&	30.772	&	27.885	&	3.529	&	3.527 \\
    7-carbophene    	&	57.952	&	0.439	&	40.205	&	35.015	&	37.220	&	34.298	&	3.530	&	3.526 \\
    8-carbophene    	&	68.455	&	0.439	&	46.886	&	41.872	&	44.069	&	41.171	&	3.530	&	3.525 \\
    9-carbophene    	&	78.960	&	0.439	&	53.565	&	48.374	&	50.580	&	47.566	&	3.530	&	3.524 \\
    10-carbophene    	&	89.465	&	0.439	&	60.255	&	55.200	&	56.300	&	54.474	&	3.531	&	3.523 \\
\br
\end{tabular}
\end{scriptsize}
\label{table:energylengths}
\end{table*}

Table S1 and  Figure S1 in the Supplemental Information contains a listing of the bond lengths found in each carbophene. They demonstrate that the degree of bond length alternation (BLA) increases for 6-member rings at the vertices as the value of N increases, while tends to decrease for the other 6-member rings.  Further, the BLA grows in the 4-member rings as N increases. Thus, the local antiaromaticity of the vertices decreases while the antiaromaticity of the other rings increase.  These observations suggest that the overall aromaticity of the carbophenes is low.


Several measures were used to determine the carbophene pore sizes; the results of which are given in Table \ref{table:energylengths}.  The simplest measure, HH, is the Euclidean distance between hydrogen atoms on opposite sides of a pore.  Next, instead of the H atoms, the corresponding carbon atoms were used in computing the Euclidean distance, CC. The carbon-carbon Euclidean distance is considered because it's been shown in other porous carbon materials that at times hydrogen atoms are bent out of the plane to allow the passage of a molecule.\cite{Raju20183969}  A simple model where the pore size, S, is equivalent to the diameter of the largest sphere that can fit through the pore without overlapping an electron shell of any of the atoms was also used. In finding S, the electron shells are assumed to have the empirical radii given by Slater.\cite{Slater19643199} The placement of the center of the sphere was determined by finding the midpoint of the line connecting the hydrogen atoms in the HH Euclidean distances.

\begin{figure}
\centering
\includegraphics[clip,width=6cm, keepaspectratio]{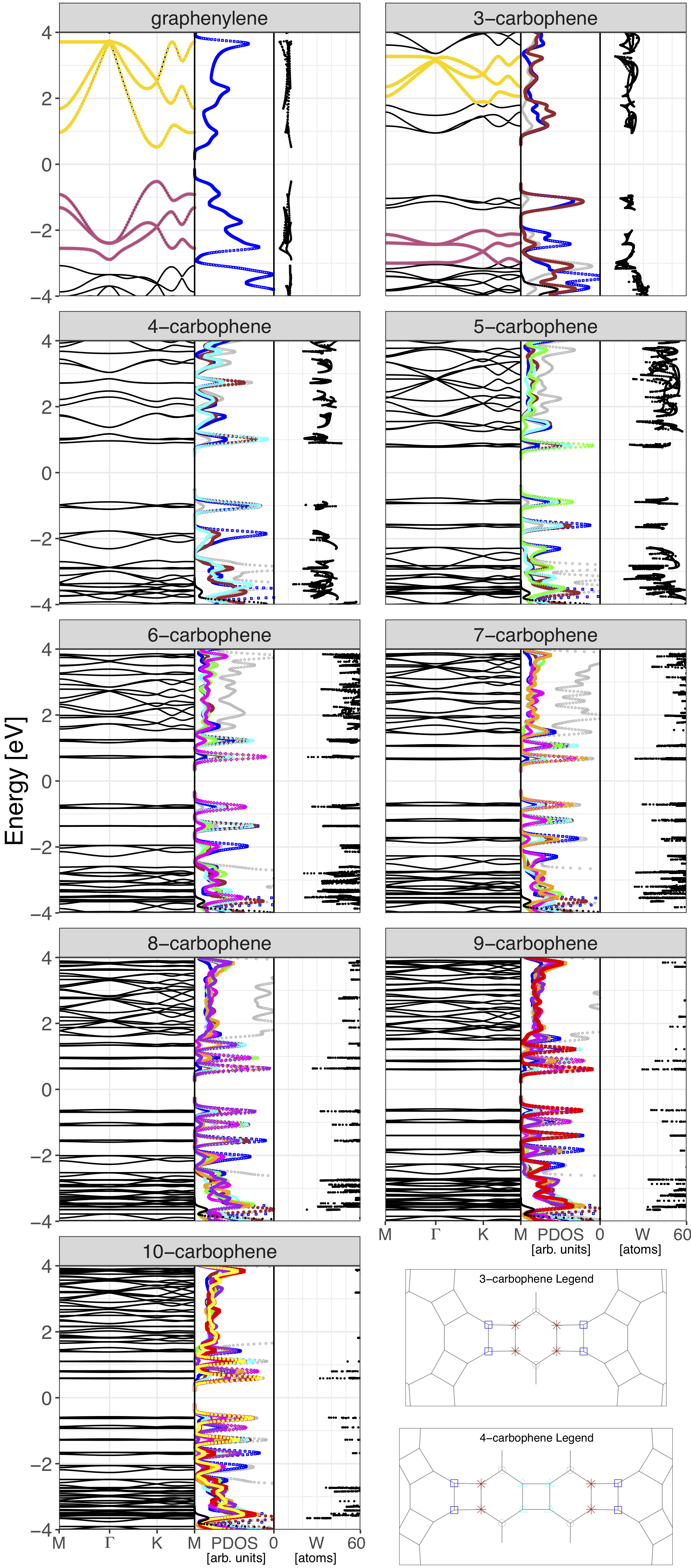}
\caption{Band dispersion, PDOS, and W for each state at each kpoint in the band dispersion. PDOS legends for 3-carbophene and 4-carbophene is on the bottom right (PDOS legends for the other structures are in the Supplemental Information).}
\label{fig:banddos}
\end{figure}

The band dispersion, the PDOS, and the number of accessible atoms, $W$, of graphenylene and each carbophene are presented in Figure \ref{fig:banddos}.  Graphenylene has well-defined conduction and valence bands. To guide the eye, the top three graphenylene valence bands are highlighted rose, and the bottom three conduction bands of graphenylene are highlighted gold.  In 3-carbophene the highlighted graphenylene bands appear again, but they are now respectively pushed away from the Fermi level and compressed by three energy bands. While these new valence and conduction bands appear to have narrow dispersion, the large unit cell leads to an electron effective mass of 0.026 $m_0$ (-0.028 $m_0$ effective hole mass), where $m_0$ is the free electron rest mass.  With 4-carbophene three new valence bands and three new conduction bands appear, further compressing the graphenylene conduction and valence bands, while also moving the 3-carbophene valence and conduction bands away from the Fermi level.  The compression of the bands increases the magnitude of the effective masses in 4-carbophene to 0.060 $m_0$ for the electron and -0.076 $m_0$ for the hole.  This pattern of adding more bands while compressing and displacing bands continues with increasing N.  As the value of N increases the valence and conduction bands narrow but how this effects the effective masses cannot be determined due to the non-parabolic nature of the bands around the high symmetry kpoint. Further analysis of the effective masses is given in the Supplemental Information.

Before computing the PDOS, the carbophenes were partitioned into groups according to where the atoms are within a phenylene segment.  The bottom right corner of Figure \ref{fig:banddos} contains representations of the 3-phenylene and 4-phenylene segments respectively within 3-carbophene and 4-carbophene, where the atom positions are marked by symbols denoting how the atoms are labeled in the PDOS; the atoms in the other carbophenes are similarly partitioned. Because the s-orbital contributions are nearly zero in the valence (conduction) bands only the p-orbitals are shown in the PDOS.  In the valence (conduction) bands, the atoms that play the most significant role are the pairs of atoms that sit at the junction of a 4-member ring and a 6-member ring.  In 3-carbophene the atoms labeled with the brown stars contribute the most to the DOS of the valence (conduction) bands, with the blue square atoms playing a slightly lesser role, the gray circle atoms play only a minor role, while the hydrogen plays almost no part in the energy range. In 4-carbophene, the addition of the cyan triangle atoms, pushes the blue square and brown star atoms further away from the center of the phenylene segment, further compressing the graphenylene style conduction and valence bands and pushing the bands further away from the Fermi level, while also compressing and displacing the 3-carbophene bands above and below the Fermi level.  Continuing to carbophenes with larger N values, the atoms nearest the center of the phenylene segments are strongly associated with bands nearest the Fermi level, with the other atoms playing smaller roles as their distance from the center of the phenylene segment increases.  Also, in continuing to carbophenes with larger N values, the bands that had been closer to the Fermi level are moved away from it.

Given an electron in state $\nu$ with wavefunction $\psi(\nu)$ one method for computing the probability of finding the electron on atom $i$ is the Mulliken population, $p_i(\nu)$. The Mulliken population of a state $\nu$ is normalized by summing over all of the atoms in a system, $\sum_{i}p_{i}(\nu) = 1$.  Using the Mulliken population with the definition of information entropy $S(\nu) = - \sum_{i} p_{i}(\nu) \ln p_{i}(\nu)$ it is possible to determine the number of atoms that are accessible to an electron in state $\psi(\nu)$ using the Boltzmann equation $W(\nu) = e^{S(\nu)}$.\cite{Junkermeier2008accessibleatoms}  Thus, given a state $\phi(\mu)$ with equal probabilities over $M$ atoms, the number of accessible atoms is $W(\mu) = M$.\cite{Lewisjp026772u}  Each graph of Figure \ref{fig:banddos}, includes the value of $W$ for each state at each kpoint.  If the valence (conduction) bands were due to defect states, one would expect to find each state to only be associated with a few atoms, yet the states are associated with about a dozen atoms in 3-carbophene, and up to 30 or more atoms for 10-carbophene.  Graphene nanoribbon (GNR) edge states are similar to defect states but extended over more atoms because of the regularity of the position of the (often passivated) dangling bond states in the GNR. The carbophene states in question are not similar to the GNR edge states because whereas the edge states are focused on the (un)passivated carbon atom, in carbophene valence (conduction) bands are centered on the cyclobutene atoms.

\begin{figure}
\centering
\includegraphics[clip,width=6cm, keepaspectratio]{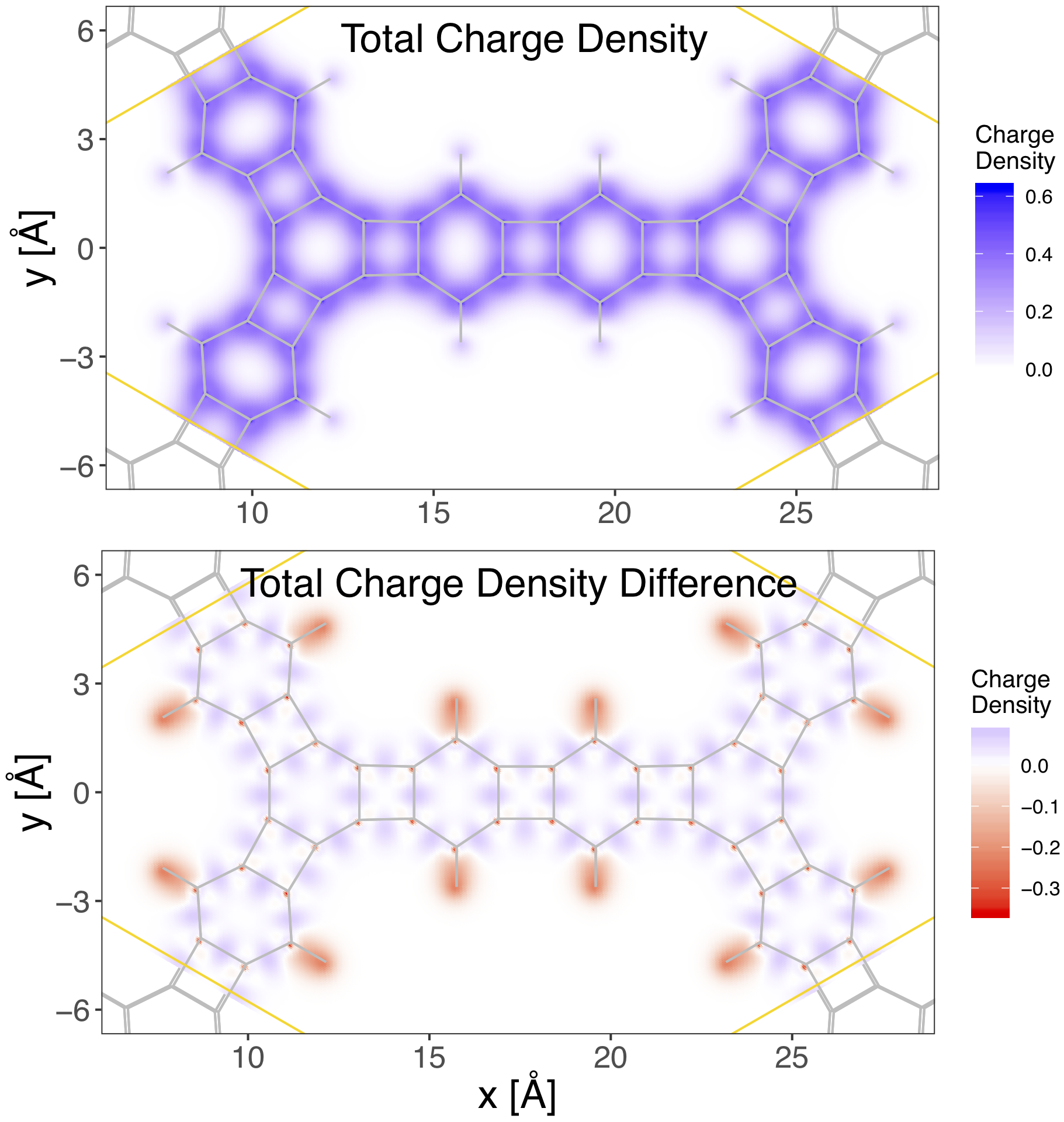}
\caption{2D projections of the total charge density and the charge density difference in 4-carbophene. The plotted map refers to the integration of the charge density along the direction perpendicular to the basal plane.  The grey lines indicated the bonds between atoms in 4-carbophene, while the gold lines represent portions of the lattice vectors defining the cell structure.}
\label{fig:totalcharge}
\end{figure}

Figure \ref{fig:totalcharge} shows the 2D projections of the total charge density and the calculated charge density difference between the value that would be found considering a superposition of neutral atoms and the charges found after a self-consistent charge (SCC) calculation. The presented results were integrated along the direction perpendicular to the basal plane for each point. The charge density difference results suggest a tendency of the electrons of the system to transfer from the hydrogen atoms to regions between adjacent carbon atoms.  The total charge density results indicate that the electrostatic effect is mainly due to the cyclobutene and cyclohexatriene units with only a small contribution from the H atoms.  That the hydrogen plays only a minor role is to be expected because this result is analogous to XRD structure analysis which seldom observes hydrogen atoms.  As is further expected, the pore region does not contribute to the electrostatic effect of the material. The geometrical configuration of the electrostatic pores in these materials will be relevant in their use in molecular docking and molecular separation.  One notices that the total charge density is transferred from the H atoms (and the C-H bond) to the bonding regions between carbons.  Mulliken charge analysis is discussed in the Supplemental Information.

\begin{figure}
\centering
\includegraphics[clip,width=6cm, keepaspectratio]{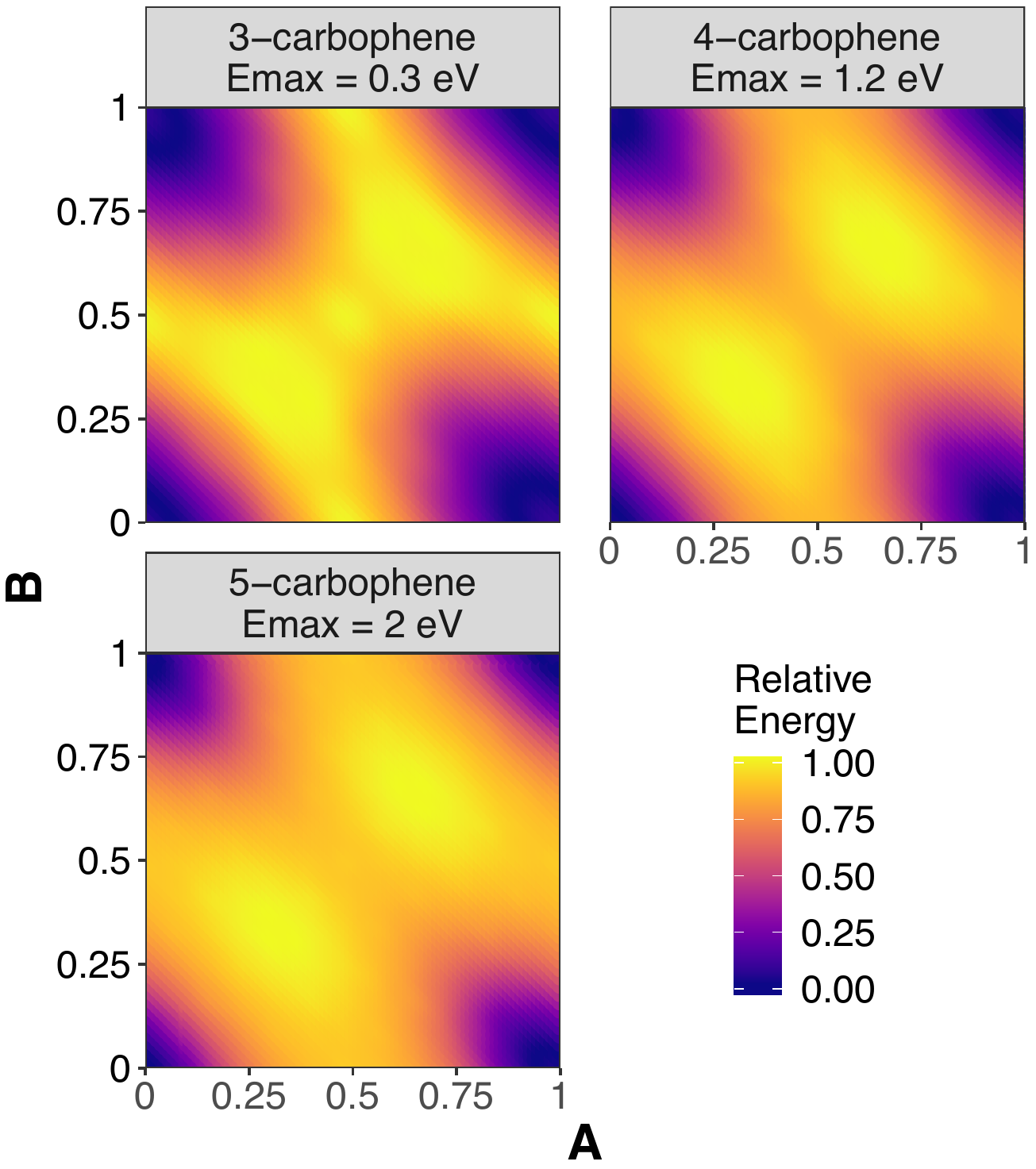}
\caption{Relative total energies with respect to the AB-plane components of displacements. The relative total energy at any point is Emax times the Relative Energy.}
\label{fig:nABbilayers}
\end{figure}

We determined the energies of bilayer 3-carbophene, as well as bilayer 4-carbophene and bilayer 5-carbophene, 
as one layer was displaced with respect to another.  A full description of the process is discussed in Reference \cite{junkermeier2019bilayers} with a brief introduction here.  The displacement vector defines the relative positions,
\begin{equation}\label{eqn:D}
\vec{\textit{D}}(a,b,h) = a \hat{\textit{A}} + b \hat{\textit{B}} + h \hat{z},
\end{equation}
where $\hat{\textit{A}}$ and $\hat{\textit{B}}$ are the in-plane lattice vectors of the unit cell, $a$ and $b$ form the ordered pair $[a,b]$ against which the energies are graphed in Figure \ref{fig:nABbilayers}. The interlayer spacing, $h_{opt}(a,b)$, is not assumed to remain constant but is found through a fitting procedure after determining the total energy at a set of interlayer spacing
\begin{equation}\label{eqn:h}
h \in \{3.3, 3.325, 3.35, ..., 3.6\},
\end{equation}
for each ordered pair $[a,b]$.  This method correctly predicted that the ground state of bilayer graphene is in the AB stacking, while predicting that bilayer graphenylene has ground state configurations that are near the in-plane displacements $[a,b] = \{[0.13, 0.43],\allowbreak [0.43, 0.13],\allowbreak  [0.57, 0.87],\allowbreak [0.87, 0.57]\}$.\cite{junkermeier2019bilayers}  Figure \ref{fig:nABbilayers} demonstrates that AA stacking is the ground state configuration for the N-carbophenes studied. When moving away from the AA stacking configuration the larger systems have a larger energy hill to climb.  If these structures stay relatively near the AA stacking at standard temperature and pressure, and if this holds for many-layered systems (not just bilayers), then many-layer carbophens may be a reliable method for creating arrays of nanochannels.\cite{Masuda1997, Li201717523}

Because of the high computational cost associated with the high throughput analysis performed in obtaining the results in Figure \ref{fig:nABbilayers} the full bilayer analysis discussed above was not performed for the larger carbophenes (N $>$ 5).  It is assumed that the AA stacking preference, as found in Figure \ref{fig:nABbilayers}, will hold for N $>$ 5.  Thus, only the interlayer spacing of the AA stacking configuration of the larger carbophenes were calculated. The interlayer spacings of bulk systems were also computed in the AA stacking.  Table \ref{table:energylengths} gives the AA stacking interlayer spacing for both the bilayer and bulk of each carbophene.  Each of the interlayer spacings listed is $~ 0.1$ \AA\, larger than the value previously found for graphenylene.\cite{junkermeier2019bilayers}

In their paper announcing the synthesis of graphenylene, Du \textit{et al.} noted  adsorbed oxygen that they couldn't account for if graphenylene was synthesized and conceded that they may have produced 3-carbophene with oxygen functional groups on the bonding sites.
Using angle resolved XRD on their sample Du \textit{et al.} give a peak diffraction angle of $2 \theta \approx \ang{23} $, which the Bragg equation turns into an interlayer spacing of $3.87$ \AA. Reanalysis of Du's XRD data places the peak diffraction angle at $2 \theta = \ang{24.17}$, or possibly to $\ang{25.22}$.\cite{junkermeier2019bilayerdata} The change in $2\theta$ shrinks the interlayer spacing from $3.87$ \AA\, to $3.68$ \AA, and possibly to $3.53$ \AA. Whatever the interlayer spacing, the fact that carbophenes have an interlayer spacing that is larger than the value found for graphenylene, lends credence to the hypothesis that Du \textit{et al.} synthesized carbophene instead of graphenylene.

\begin{figure}
\centering
\includegraphics[clip,width=6cm, keepaspectratio]{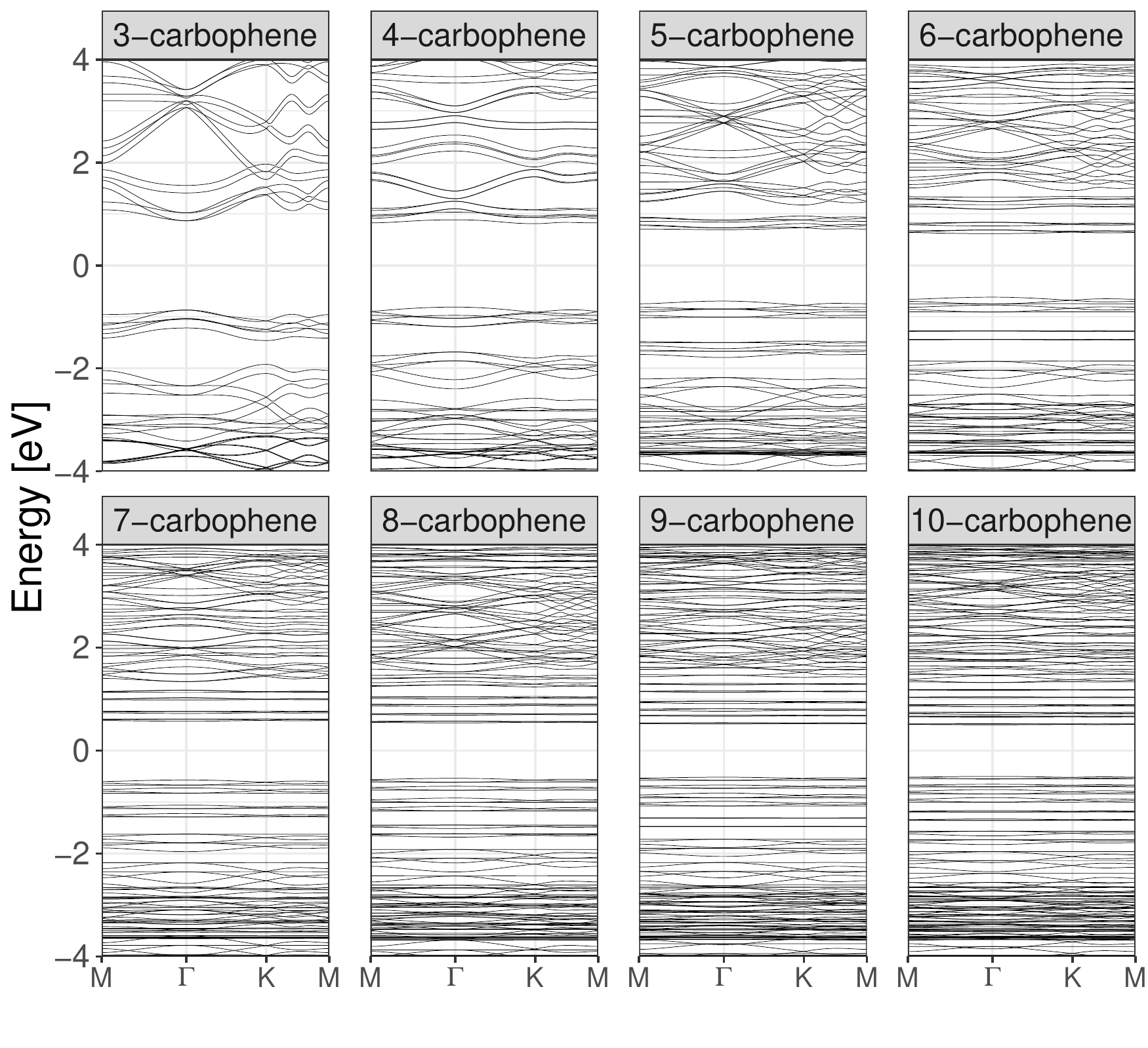}
\caption{Band structure plots evidencing the appearance of nearly degenerate bands 
for N-carbophenes with $3\le N\le10$.}
\label{fig:bilayerBAND}
\end{figure}
\begin{figure}
\centering
\includegraphics[clip,width=6cm, keepaspectratio]{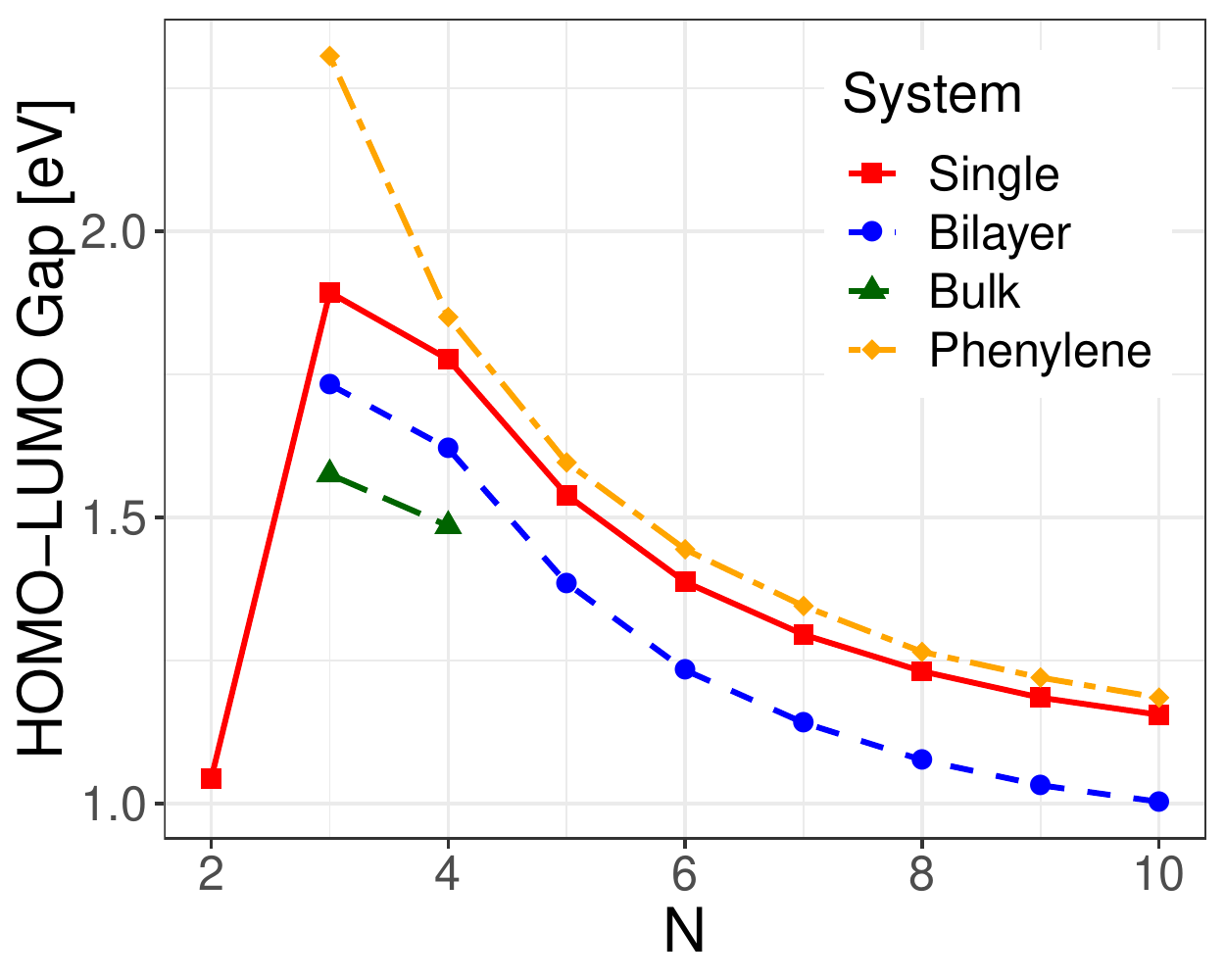}
\caption{Comparison of the HOMO-LUMO energy differences of linear \textit{N}-phenylene with band gaps of single layer, bilayer, and bulk \textit{N}-carbophene. The x-axis signifies the value of N in \textit{N}-phenylene, and in \textit{N}-carbophene.}
\label{fig:bandgaps}
\end{figure}

As in bilayer graphene (BLG), and ultimately graphite, the carbophene layers interact weakly. The weakly interacting layers in BLG are responsible for the formation of nearly degenerate bands.\cite{AOKI2007123} Nearly degenerate bands appear in multilayer carbophenes as well; as is demonstrated in Figures \ref{fig:bilayerBAND}.  The near degeneracy of the bands widens the valence (conduction) band energy regions narrowing the energy gap between valence (conduction) bands and decreasing the band gap values with the occurrence of more layers.  Figure \ref{fig:bandgaps} presents the molecular HOMO-LUMO energy difference or band gap in extended systems with moving from the linear \textit{N}-phenylenes to the single layer carbophenes, bilayer carbophenes, and bulk carbophenes.

\section{Conclusion}

Using tight-binding density functional theory, we show that carbophenes have formation energies per carbon atom similar to that of graphenylene.
The similarity in formation energies between graphenylene and carbophene suggests that when trying to synthesize graphenylene, a carbophene may be synthesized, or vice versa. The similar formation energies explain why the first reported synthesis of graphenylene reported that 3-carbophene might have been synthesized instead.\cite{du1740796} Using high-throughput computing, we found that AA stacking is the most stable configuration of carbophene bilayers. After assuming that AA stacking is the ground state configuration for bulk carbophene, the interlayer spacing of bulk carbophenes were determined. This work compared the theoretical interlayer separations of multilayer graphenylenes and multilayer carbophenes with the experimental value found by Du \textit{et al.}.   We concluded that a carbophene would be the more likely candidate for what Du \textit{et al.} produced.\cite{du1740796, junkermeier2019bilayers} Du \textit{et al.} reported that their synthesized material contained oxygen, thus to better fit this condition, future work should replace hydrogen atoms with oxygen-containing functional groups.


\section*{Supplementary material}

Supplementary material for this article is available online.

\section*{Acknowledgements}
RP acknowledges FAPESP for financial support through project
$\#$2018/03961-5 and CNPq for grant $\#$310369/2017-7.  This research was supported by resources supplied by the Center for Scientific Computing (NCC/GridUNESP) of the S\~ao Paulo State University (UNESP).

\section*{{CRediT} authorship contribution statement}

Chad E. Junkermeier: Conceptualization, Data curation, Formal analysis, Investigation, Methodology, Project administration, Resources, Software, Validation, Visualization, Writing - original draft, Writing - review \& editing.
Jay Paul Luben: Investigation, Formal analysis, Writing - review \& editing.
Jordan Dalessandro: Formal analysis.
Ricardo Paupitz: Investigation, Resources, Visualization, Writing - review \& editing.
Turner S. Reed: Formal analysis.

\section*{conflict of interest}
The authors declare no conflict of interest.


\bibliographystyle{iopart-num.bst}
\bibliography{carbophene}



\end{document}